\documentclass[12pt,a4paper]{article}

\usepackage{feynmp,amsbsy,latexsym,cite}

\newcommand{\no}{\noindent}

\newcommand{\ad}{{a^{\dag}}}
\newcommand{\cd}{\cdot}
\newcommand{\ecd}{{\cdot}}

\newcommand{\pa}{\partial}
\newcommand{\G}{{\cal G}}

\newcommand{\ee}{\mathrm{e}}

\newcommand{\free}{{\mathrm{f}}}

\newcommand{\Afree}{A^\free}
\newcommand{\soft}{{\mathrm{soft}}}

\newcommand{\as}{{\mathrm{as}}}
\newcommand{\Haas}{{ H}_{\mathrm{int}}^{\mathrm \as}}

\newcommand{\Aas}{A^\as}
\newcommand{\xb}{{\boldsymbol x}}
\newcommand{\yb}{{\boldsymbol y}}

\newcommand{\pb}{{\boldsymbol p}}

\newcommand{\vb}{{\boldsymbol v}}
\newcommand{\intx}{\int\!d^3x\;}

\newcommand{\intp}{\int\!\frac{d^3p}{(2\pi)^3}\;}

\newcommand{\intk}{\int\!\frac{d^3k}{(2\pi)^3}\;}

\newcommand{\intks}{\int\limits_{\soft}\!\frac{d^3k}{(2\pi)^3}\;}

\begin{document}

\begin{titlepage}
\rightline{UAB-FT-00-7}
\rightline{PLY-MS-00-7}

\vskip19truemm
\begin{center}{\Large{\textbf{Electrons and Photons: Fact not
 Fiction}}}\\ [12truemm]
\textsc{Emili Bagan}\footnote{email: bagan@ifae.es}\\
[5truemm] \textit{Dept.~Fisica Te\`orica \&\ IFAE\\
Edifici Cn\\
Universitat Aut\`onoma de Barcelona\\E-08193 Bellaterra  (Barcelona)\\
Spain\\[5truemm]}
\textsc{Martin Lavelle}\footnote{email: mlavelle@plymouth.ac.uk} and
\textsc{David McMullan}\footnote{email: dmcmullan@plymouth.ac.uk}\\
[5truemm] \textit{School of Mathematics and Statistics\\ The
University of Plymouth\\ Plymouth, PL4 8AA\\ UK} \end{center}

\bigskip\bigskip\bigskip
\begin{quote}
\textbf{Abstract:} The particle Fock space of the matter fields in
QED can be constructed using the free creation and annihilation
operators. However, these particle operators  are not, even at
asymptotically large times, the modes of the matter fields that
enter the QED Lagrangian. In this letter we construct the fields
which do recover such particle modes at large times. We are thus
able to demonstrate for the first time that, contrary to
statements found in the literature, a relativistic description of
charged particles in QED exists.
\end{quote}

\end{titlepage}

\setlength{\parskip}{1.5ex plus 0.5ex minus 0.5ex}

\noindent \no \textbf{Introduction}

\smallskip

\no The conclusions of this paper, that electrons are particles,
will not surprise our experimental colleagues. What may come as a
surprise to them is the fact that the hitherto accepted
wisdom~\cite{kulish:1970,Zwanziger:1976vv,Frohlich:1979bf,Buchholz:1991dx}
in the theoretical community was that there does not exist any
relativistic description of the electron as a particle!

The root of the problem~\cite{Dollard:1964} lies in the
masslessness of the photon. This generates long range interactions
and it is well known that these fall off so slowly that they
cannot be neglected even for widely separated charges a long time
before or after scattering processes. This is seen in $S$-matrix
calculations as the lack of a pole for fermionic external legs.

This lack of a particle interpretation of an electron presents a
radical departure from the usual view of particle physics which we
find highly unsatisfactory. We want to show that it is also unnecessary.
We note that an understanding of how particles can arise in gauge theories
is important to improve our insight into the physical structures of
gauge theories, to help with QCD phenomenology and it may well have spin
offs for the vexed question of how to describe unstable particles.

In the context of QED this lack of any particle language has not
hindered progress. Our knowledge of the classical limit of QED
acts as a guide to the extraction of physical predictions from
suitably defined cross-sections via the Bloch-Nordsieck framework.
In theories such as QCD such  intuition is still greatly lacking.
In particular we do not understand hadronisation which relies on
coloured particles metamorphosing into jets. The route which leads
from partons to effective quarks and glue is essentially uncharted.

In this letter we will show that a relativistic particle description  of the
electron (or any other charged particle) immediately follows once a
correct physical identification of the electron has been made. Having
first recalled the standard statement of the
problem~\cite{kulish:1970}, we will show that the effects which
prevent the identification of a charged particle structure in QED
disappear if the right fields are used. We stress that this
identification only holds in the asymptotic region a long time before
or after scattering occurs. We then demonstrate that a
particle structure is also asymptotically
present for photons in full QED. Finally we discuss how these results
have been verified in perturbative calculations and present some
conclusions.

\bigskip

\noindent \no \textbf{The Interaction}

\smallskip

\no We have already noted that in theories like QED we have long range
interactions due to
the masslessness of the photon. It has been shown by various
authors that such interactions cannot be neglected even at large
times before or after scattering. In particular, Kulish and
Faddeev~\cite{kulish:1970} showed that the annihilation operator
of the matter fields of QED which is defined as the large time
limit of the operator
\begin{equation}\label{2bbad}
  b(q,s, t):=\intx\frac1{\sqrt{2E_{\smash{q}}}}u^{\dag s}(q)\psi(x)\ee^{iq\ecd
  x}\,,
\end{equation}
does not become just the usual free particle mode, but rather takes on
the form\footnote{There is also a distortion contribution to the non-observable
phase of $S$-matrix elements which we will ignore in this letter.}
\begin{equation}\label{2bbadr}
  b(q,s,t)=D_\soft(q,t)b(q,s)\,,
\end{equation}
where
\begin{equation}\label{2dsoft}
  D_{\soft}(q,t)=\exp\left\{-e\!\!\!\intks\frac1{2\omega_k}
  \left(
  \frac{q\cd a(k)}{q\cd k}\ee^{-itk\ecd q/E_q}-
\frac{q\cd \ad(k)}{q\cd k}\ee^{itk\ecd q/E_q}
  \right)\right\}\,,
\end{equation}
is called a \emph{distortion operator}~\cite{kulish:1970}. The
creation and annihilation operators for the photonic variables
enter into this expression and, as long as $D_{\soft}\ne1$, a
particle mode for the electron will not be recovered. Of course,
we should only expect to  regain a particle description at
asymptotic times and, for large $t$, the integral in
(\ref{2dsoft}) only receives contributions from soft photons, but
it still does not reduce to the unit operator. This important
observation has lead to the conclusion~\cite{kulish:1970} that it
is \emph{not} possible to describe the electron as a particle.

However, since the interaction does not switch off, the matter field
$\psi$ \emph{never} becomes gauge invariant and thus we should not expect to
identify it with physical particles via Equation~\ref{2bbad}!
We now construct the fields which do have a particle description at
large times.

\bigskip

\noindent \no \textbf{The Electron}

\smallskip

\no We  recall that physical fields must be invariant under the local gauge
transformations, $A_\mu(x)\to A_\mu(x) + \partial_\mu\theta(x)$ and
$\psi(x)\to \ee^{ie\theta(x)}\psi(x)$. A physical matter field must
then be given by a product of the form $h^{-1}(x)\psi(x)$ where,
under a gauge transformation, $h^{-1}(x)
\to h^{-1}(x) \ee^{-ie\theta(x)}$. We call such a field with this
gauge transformation property a dressing.

Of course there are a multiplicity of such dressing fields which
satisfy this minimal
requirement~\cite{Dirac:1955ca,Lavelle:1997ty}. To describe a
charged particle we need to further
demand~\cite{Bagan:1997kg} that the dressing
satisfies the dressing equation
\begin{equation}\label{4de}
  u\cd\pa h^{-1}(x)=-ieh^{-1}(x)u\cd A(x)\,,
\end{equation}
where $u^\mu=\gamma(\eta+v)^\mu$ is the four velocity of the charged particle,
$\eta$ is the unit time-like vector, $v=(0,\vb)$ is the velocity
and $\gamma=(1-|\vb|^2)^{-1/2}$.
It is important to note the velocity dependence here: we can only
expect to have a particle interpretation of a charge at asymptotic
times and in that regime the velocity is a well defined quantum
number. Our dressed charges will, therefore, only correspond to particles at the
appropriate point on the mass shell characterised by the velocity in
the dressing.

In QED we have been able to solve these two
requirements~\cite{Bagan:1997kg,Bagan:1999jf} and have found the
following description of a charged field moving with a given
velocity:
\begin{equation}\label{4sol}
  h^{-1}(x)\psi(x)=\ee^{-ieK(x)}\ee^{-ie\chi(x)}\psi(x)\,.
\end{equation}
The $K$ term is separately gauge invariant and contributes to the
unobservable phase. We ignore it in this letter. The $\chi$ part of the
dressing may be written as
\begin{equation}\label{4min}
  \chi(x)=\frac{\G\cd A}{\G\cd\pa}\,,
\end{equation}
where  $\G^\mu=(\eta+v)^\mu(\eta-v)\cd\pa-\pa^\mu$.

We now define an annihilation operator for the gauge invariant, dressed  field
\begin{equation}\label{2bgood}
  b(q,s,v,t):=\intx\frac1{\sqrt{2E_{\smash{q}}}}u^{\dag s}(q)\ee^{-ie\chi(x)}
  \psi(x)\ee^{iq\ecd
  x}\,.
\end{equation}
Note that this has an explicit velocity dependence coming from the
form of the dressing. To find its asymptotic form we proceed in the
same manner as Kulish and Faddeev~\cite{kulish:1970}. The
annihilation operator now picks up two distortion factors: the
original one associated with the unphysical matter field and a further
correction from the dressing. We find
\begin{equation}\label{3bexp}
  b(q,s,t,v)\to h^{-1}_\soft(q,t,v)D_\soft(q,t)b(q,s)\,,
\end{equation}
where  $D_\soft(q,t)$ is given in (\ref{2dsoft}). The distortion from
the soft part of the
dressing is
\begin{equation}\label{3ddress}
  h^{-1}_{\soft}(q,t,v)=\exp\left\{e\!\!\!\intks\frac1{2\omega_k}
  \left(
  \frac{V\cd a(k)}{V\cd k}\ee^{-itk\ecd q/E_q}-
\frac{V\cd \ad(k)}{V\cd k}\ee^{itk\ecd q/E_q}
  \right)\right\}\,,
\end{equation}
where $ V^\mu=(\eta+v)^\mu(\eta-v)\cd k-k^\mu$, which is essentially
the Fourier transform of $\G^\mu$.
Combining these  distortions we find the overall distortion factor
\begin{eqnarray}
 h^{-1}_\soft(q,t,v)D_{\soft}(q,t)&=&\exp\Bigg(e\!\!\intks\frac1{2\omega_k}
  \bigg[\bigg(
  \frac{V\cd a(k)}{V\cd k}-\frac{q\cd a(k)}{q\cd k}\bigg)\ee^{-itk\ecd q/E_q}\nonumber\\
  &&\qquad\qquad\qquad-
\bigg(\frac{V\cd \ad(k)}{V\cd k}-\frac{q\cd \ad(k)}{q\cd k}\bigg)\ee^{itk\ecd q/E_q}
  \bigg]\Bigg)\,.
\end{eqnarray}
We now note that we can write (recall that $k$ is on-shell)
\begin{eqnarray}
\frac{V^\mu}{V\cd k}-\frac{q^\mu}{q\cd k}&=& \frac{(\eta+v)^\mu(\eta-v)\cd
k-k^\mu}{(\eta+v)\cd k(\eta-v)\cd k}-\frac{q^\mu}{q\cd k}\nonumber\\
&=&\frac{(\eta+v)^\mu}{(\eta+v)\cd k}-\frac{q^\mu}{q\cd k}-\frac{k^\mu}{V\cd k}\,.
\end{eqnarray}
Hence, at the point in the mass-shell where $q^\mu=m\gamma(\eta+v)^\mu$, the
distortion operator becomes a trivial operator since the argument of
the exponential becomes
\begin{equation}
-e\intk\frac1{2\omega_k}
 \bigg(
  \frac{k\cd a(k)}{V\cd k}\ee^{-it\omega_k}
  -\frac{k\cd a^\dag(k)}{V\cd k}
  \ee^{it\omega_k}\bigg)\,,
\end{equation}
which vanishes between physical states due to the Gupta-Bleuler
subsidiary condition (it only contains unphysical photon degrees of
freedom which correspond to the Nakanishi-Lautrup $B$ field).
We thus see that the modes of
the dressed field, at the appropriate place in the mass shell, are
free particle modes at large times. There is no distortion and we
have recovered a particle picture for charged matter in QED.

\bigskip

\noindent \no \textbf{The Photon}

\smallskip

\no In the usual framework  it is assumed that the coupling
switches off asymptotically.
There are then many ways to show that the physical components of
the non-interacting vector potential are the transverse degrees of freedom (see, e.g.,
Chap.~19 of~\cite{Henneaux:1992ig}).
However, we have seen that
in
the true asymptotic domain the coupling does not vanish due to the masslessness of the
photon. This means that previous arguments are incomplete. We will
now show that, although we still have an interaction, the
physical photonic degrees of freedom do decouple and a particle
description for the photon can be recovered in full QED.

The asymptotic form of the (interaction picture) vector boson
can be readily obtained once the form of the asymptotic interaction Hamiltonian, $\Haas$, has been
identified~\cite{kulish:1970,Horan:1999ba}. Using this result we can
straightforwardly transform from the free vector boson, $\Afree$, to
the asymptotic Heisenberg field, $\Aas$,  as follows:
{\setlength\arraycolsep{2pt}
\begin{eqnarray}\label{2aas}
 \Aas_\mu(x)&=&\exp\!\!\left(i\!\!\int_{-\infty}^t\!\!\!\!\!\!
 d\tau\,\Haas(\tau)\right)\Afree_\mu(x)
 \exp\!\!\left(-i\!\!\int_{-\infty}^t\!\!\!\!\!\!d\tau\,
 \Haas(\tau)\right)\nonumber\\
&=&\Afree_\mu(x)-e\int_{-\infty}^t\!\!d\tau d^3y\,D(\tau-t,
\yb-\xb)J^\as_\mu(\tau,\yb)\,,
\end{eqnarray}}%
where the asymptotic current is given by
\begin{equation}\label{2jas}
  J^\mu_{\as}(t,\xb)=\intp\frac{p^\mu}{E_p}\rho(p)\delta^3
  \bigl(\xb-t\pb/E_p\bigr)\,.
\end{equation}
This shows, as expected, that the vector boson field is also not free at large times.
In the spirit of our discussions above, we can, though, straightforwardly recover a
particle picture for photons by showing that the second term in (\ref{2aas})
vanishes for the transverse,  physical components,
$(\delta_{ij}-\partial_{i}\partial_{j}/\nabla^2)\Aas_{j}$. This shows
that in the far field domain, where the potential takes on the
asymptotic form (\ref{2aas}), a particle description emerges even in the interacting
theory.

\bigskip

\noindent \no \textbf{Conclusions}

\smallskip

\no In this letter we have shown how a particle description for the
electron and the photon emerges from QED. The essential observation
in this construction is that a particle cannot be identified with the
raw matter field that
enters into the QED Lagrangian. Rather, a physical particle
corresponds to an appropriately dressed, gauge invariant field.
The dressing for a charged  asymptotic particle with a specific velocity
depends explicitly on that velocity. A particle description for the electron
is only recovered at that point in the mass shell corresponding to
the
velocity.

For a photon it is not  surprising that not all the components of the
vector potential are physical. What is new in our discussion of
the photon is that we have
seen how the free photon emerges in the asymptotic regime even though there
is still a residual interaction with matter for the
non-physical components.

The formal arguments presented in this letter have been checked in
a wide variety of detailed calculations. We have
shown~\cite{Bagan:1998kg,Bagan:1999jk}
for both scalar and fermionic QED that the on-shell propagator and
other Green's functions of the dressed matter fields have, to all
orders in perturbation theory, a good
pole structure. This is in contrast to the Lagrangian matter fields whose
on-shell Green's functions are plagued by infra-red singularities.
This means~\cite{Bagan:1999jf} that for our fields the $S$-matrix
elements can be constructed via the traditional LSZ-formalism.

We should also point out that these charged fields have good
ultra-violet behaviour~\cite{Bagan:1997su,Bagan:1999jk}.
Multiplicative renormalisation is possible and the dressed field
operators do
not mix under renormalisation~\cite{Bagan:1999jk}. In addition,
there is an ultra-violet logarithm in scattering
processes~\cite{Bagan:1999jk} which is just the universal
Isgur-Wise (or equivalently the Wilson line kink) renormalisation
constant. This logarithm structure is just the one that appears
in the Bloch-Nordsieck formalism.

This programme is currently being extended to the non-abelian
domain where the theoretical problems are more
severe~\cite{Ciafaloni:1989vs,buchholz:1996} and the
experimental situation in identifying physical quarks and gluons,
e.g., in jets, is much more subtle.
The essential new ingredient found there that further obstructs a
particle interpretation is the gluonic self-interaction. Such
massless charges spawn a new class of asymptotic  dynamics which
is no longer spin independent and generates collinear
singularities. This can be studied in massless QED and early
indications~\cite{Horan:1998im} are that the appropriate dressed
massless charges are indeed free. Furthermore, we have already
seen~\cite{Lavelle:1998dv} that the minimal, non-abelian component
of the dressing is responsible for the anti-screening interaction
that drives asymptotic freedom in both three and four
dimensional QCD. These results give us a great deal of confidence
that we can regain a particle picture of quarks and gluons in the
pre-hadronisation regime.

\bigskip

\no\textbf{Acknowledgements:} This work was supported by the
British Council/Spanish Education Ministry \textit{Acciones
Integradas} grant no.\  1801/HB1997-0141. We thank Robin Horan,
Tom Steele, Shogo Tanimura and Izumi Tsutsui for discussions.


\end{document}